%% file: main.tex
\documentclass{llncs}

\usepackage{etex}

\usepackage{wrapfig}
\usepackage{times}
\usepackage{proof}
\usepackage{amssymb}
\usepackage{amsmath}
\usepackage{url}
\usepackage{color}
\usepackage{multicol}
\usepackage{multirow}
\usepackage{wasysym}
\usepackage{stmaryrd}
\usepackage{algorithmicx}
\usepackage[noend]{algpseudocode}
\usepackage{algorithm}
\usepackage{pgfplots}
\usepackage{multicol}

\newcommand{\Vampire}{{Vampire}}
\newcommand{\vampire}{\Vampire}

\newcommand{\leaveout}[1]{}

\usepackage{adjustbox}
\usepackage{array}
\newcolumntype{R}[2]{%
    >{\adjustbox{angle=#1,lap=\width-(#2)}\bgroup}%
    l%
    <{\egroup}%
}

\title{Testing a Saturation-Based Theorem Prover: Experiences and Challenges (Extended Version)
\thanks{
This work was supported by EPSRC Grant EP/K032674/1. 
Martin Suda and Andrei Voronkov were partially supported by ERC Starting Grant 2014 SYMCAR 639270. 
Martin Suda was also partially supported by the Austrian research projects FWF S11403-N23 and S11409-N23.
Andrei Voronkov was also partially supported by the Wallenberg Academy Fellowship 2014 - TheProSE.} }
\author{
 Giles Reger\inst{1} \and 
 Martin Suda\inst{2} \and 
 Andrei Voronkov\inst{1,3,4} 
}
\institute{
University of Manchester, Manchester, UK \and
TU Wien, Vienna, Austria \and
Chalmers University of Technology, Gothenburg, Sweden \and
EasyChair
}

\begin{document}
\maketitle
\begin{abstract}
This paper attempts to address the question of how best to assure the correctness of saturation-based automated theorem provers 
using our experience developing the theorem prover \vampire{}. 
We describe the techniques we currently employ to ensure that Vampire is correct 
and use this to motivate future challenges that need to be addressed 
to make this process more straightforward and to achieve better correctness guarantees. 
\end{abstract}

\input{introduction}
\input{correct}

\input{random}
\input{experience}
\input{proofcheck}
\input{challenges}
\input{conclusion}

\bibliography{bib}
\bibliographystyle{cs-abbrv++}

\end{document}

%% file: introduction.tex
\section{Introduction}\label{sec:introduction}

This paper considers the problem of checking that a saturation-based automated theorem prover is \emph{correct}. 
We consider this question within the context of the \vampire{} theorem prover \cite{KovacsVoronkov:CAV:Vampire:2013},
but many of our discussions generalise to similar theorem provers such as E \cite{DBLP:journals/aicom/Schulz02},
SPASS \cite{DBLP:conf/cade/WeidenbachDFKSW09},
and iProver \cite{DBLP:conf/cade/Korovin08}. 
We discuss what we mean by this correctness, give examples of bugs that we have encountered while developing Vampire,
describe how we detect such bugs and, as our main contribution, outline the challenges that need to be addressed.

It should not be necessary to motivate the general need to ensure that any piece of software is correct.
However, the cost of incorrect software is not uniform and here we briefly motivate why ensuring that theorem provers, such as \vampire{}, are correct is of significant importance. Theorem provers are often used as \emph{black boxes} in other techniques (e.g. program verification) and those techniques rely on the results of the theorem prover for the correctness of their own results. 
Another area that makes use of automated theorem provers is the application of so-called \emph{hammers} 
\cite{IWIL2010:Three_years_of_experience_with_Sledgehammer_a_Practical_Link_Between_Automatic_and_Interactive_Theorem_Provers1,DBLP:journals/mics/KaliszykU15}
in interactive theorem proving. These combinations usually provide functionality to reconstruct the proofs of the automated theorem provers using their own trusted kernels, although they also offer users the option to skip such steps.

It is clear that correctness is important here, so how are we doing? Most theorem provers seem to be generally correct. However, cases of unsoundness are not uncommon. In SMT-COMP 2016 there were 603 conflicts (solvers returning different results) on 73 benchmarks. This turned out to be three solvers giving incorrect results for various reasons.\footnote{See \url{http://smtcomp.sourceforge.net/2016/}.}
In the CASC competition \cite{DBLP:journals/aim/Sutcliffe16}, there is a period of testing where soundness is checked and resolved, and there have been a number of solvers later disqualified from the competition due to unsoundness. 
In our experience, adding a new feature to a theorem prover is a highly complex task and it is very easy to introduce unsoundness,
or general incorrectness, especially in areas of the code that are encountered during proof search very infrequently.

The paper makes the following contributions:
\begin{itemize}
	\item An overview of what we mean by correctness with respect to saturation-based theorem provers (Section~\ref{sec:correctness}).
	\item A description of the approach we take to bug finding (Section~\ref{sec:random}) that involves sampling the input space of proof search heuristics and problems.
	\item A sample of real world bugs that we have found in \vampire{} that demonstrates the complexity of this problem (Section~\ref{sec:bugs}).
	\item A presentation of \emph{proof checking} approaches that can help ensure the correctness of proofs (Section~\ref{sec:proofcheck}) and a reflection on what cannot currently be checked.
	\item We propose a set of challenges (Section~\ref{sec:challenges}) that need to be addressed to produce a better solution to this problem in general. 
\end{itemize}

The aim of this paper is to provide sufficient context to explain the challenges outlined in Section~\ref{sec:challenges}. Addressing these challenges is part of our current ongoing research.

%% file: correct.tex
\section{What Does Correctness Mean for Us?}\label{sec:correctness}

We separate four different ways in which \vampire{} can be incorrect:
\begin{itemize}
	\item Incorrect Result
	\item Program Crash
	\item Assertion Violation
	\item Performance Bug
\end{itemize}
We will now briefly discuss what we mean by each class of incorrect behaviour. 
The most damaging (and interesting) of these is an incorrect result and we discuss this first.

\subsection{Incorrect Result}

To understand what a correct and incorrect result mean to \vampire{} we need to introduce some of the theoretical foundations of the underlying technique. We note that the approach used by \vampire{} is the same as that taken by other first-order theorem provers,
so these discussions, and the challenges outlined in Section~\ref{sec:challenges}, generalise beyond Vampire.

Vampire accepts problems (formulas) in the form 
\begin{equation}\label{eq0}
(\mathit{Premise}_1 \wedge \ldots \wedge \mathit{Premise}_n) \rightarrow \mathit{Conjecture}
\end{equation}
and can give one of three answers:
\begin{itemize}
	\item \emph{Theorem}, if the problem is true in all models,
	\item \emph{Non-Theorem}, if there are models in which \eqref{eq0} is false, and 
	\item \emph{Unknown}, if \vampire{} cannot deduce one of the previous answers.
\end{itemize}
Providing one of the first two results when that result does not hold is clearly incorrect. 
Providing \emph{Unknown} as the result is clearly incorrect in the sense that there is a known answer, 
but, due to the undecidability of first-order-logic and the general hardness of the problem,
it is often unavoidable. 
However, as discussed below, we should understand the different ways in which \emph{Unknown} as a result can be produced. Note that \emph{Unknown} will be returned if Vampire exceeds either the time or memory allotted to it.

More specifically, \vampire{} is a refutational theorem prover;
it establishes the \emph{validity} of problems in the form~\eqref{eq0} by detecting \emph{unsatisfiability} of its negation:
\begin{equation}\label{eq1}
\mathit{Premise}_1 \wedge \ldots \wedge \mathit{Premise}_n \wedge \neg \mathit{Conjecture}.
\end{equation}
This works by translating~\eqref{eq1} into a set of \emph{clauses} ${\cal S}$
and adding consequences of ${\cal S}$ until the contradiction $\mathit{false}$ is derived 
or all possible consequences have been added.
This process is called \emph{saturation} and may not terminate in general for a satisfiable set ${\cal S}$.

If \vampire{} derives a contradiction then it has shown that the problem \eqref{eq0} is \emph{valid}, i.e. a theorem. 
Deriving a contradiction when the problem in \eqref{eq0} is not valid is \emph{unsound} and an \emph{incorrect result}.

If \vampire{} fails to derive a contradiction and \emph{saturates} the set ${\cal S}$ in finitely many steps 
then there is a result \cite{BachmairGanzinger:HandbookAR:resolution:2001} 
telling us that under certain conditions we can conclude that $\mathit{false}$ cannot be a consequence of ${\cal S}$ and therefore problem~\eqref{eq0} is a non-theorem. These conditions capture the \emph{completeness} of the underlying inference system and generally require that all possible \emph{non-redundant} inferences have been performed. 

However, there are many things that \vampire{} does to heuristically improve proof search that break the completeness conditions. For example:
\begin{itemize}
	\item Certain well-performing selection functions \cite{selections} might prevent inferences that need to be performed for completeness conditions to hold.
	\item Some useful preprocessing steps explicitly remove some of the premises \cite{DBLP:conf/cade/HoderV11}.
	\item Strategies such as the \emph{limited resource strategy} \cite{RiazanovVoronkov:JSC:LRS:2003}
	 might at some point choose to delete a clause from the search space.
	\item Reasoning with theories such as arithmetic is in general undecidable and the above conditions fail to hold.
\end{itemize}	
If the completeness conditions do not hold then upon saturation the result is unknown. 
Sometimes it is easy to detect when the completeness conditions hold, sometimes it is non-trivial and sometimes they are erroneously broken. In this last case (when we think the conditions hold but they do not) this will lead to incorrectly reporting non-theorem 
i.e. this \emph{completeness issue} is another kind of \emph{incorrect result}.

To ensure the requirement that all possible non-redundant inferences will in the end be performed,
we impose certain \emph{fairness} criteria on the saturation process. More concretely,
we require that no such inference is postponed indefinitely. 
Notice that this is by nature a tricky condition to deal with 
as it cannot be seen to have been violated after finitely many steps while the prover is running.
And since, due to the semi-decidability of first-order logic, there is no upper 
bound on the length of the computation required to derive $\mathit{false}$,
a non-fair implementation might in certain cases never be able to return \emph{Theorem},
even if it is the correct answer and instead keep computing indefinitely.
Thus, this \emph{fairness issue} does not lead to an incorrect result per se,
but rather just negatively influences performance. As such it may be extremely hard
to detect and deal with.

\subsection{Program Crash}

A program crash is where \vampire{} terminates unexpectedly, usually due to one of the following:
\begin{itemize}
	\item Unhandled Exception
	\item Floating Point Error (SIGFPE)
	\item Segmentation Fault (SIGSEG)
\end{itemize}

Unhandled exceptions are bugs as we should handle them. 
In general, \vampire{} handles all known classes of exceptions at the top level,
but we have recently had issues with integrated tools (MiniSAT and Z3) producing exceptions that we did not handle. Floating point errors and segmentation faults are typical software bugs that should be detected and removed. 
Later we discuss the potential dangers of memory related bugs.

\subsection{Assertion Violation}

\vampire{} is developed defensively with frequent use of \emph{assertions}. For example, these are inserted wherever a function makes some assumptions about its input or the results of a nested function call,
and wherever we believe a certain line to be unreachable. 
\vampire{} consists of roughly 194,000 lines of C++ code with roughly 2,500 assertions,
meaning that there is roughly one assertion per 77 lines. The majority of potential errors 
are detected early as assertion violations.

\subsection{Performance Bug}

This is not a bug in the sense of the above bugs i.e. \vampire{} still returns the correct result without early termination. However, we include this category of bugs as it is something we are interested in and try to detect.
A performance bug is where something should perform well, but due to a mistake or error it does not.
An easy case to detect is where a heuristic previously performed well but after implementing an orthogonal heuristic it does not.
A very difficult case to detect is where a new heuristic is added but implemented incorrectly and therefore does not perform as it should.
There is little hope of detecting these cases without manual inspection.

%% file: random.tex
\section{Finding Bugs} \label{sec:random}

In this section we briefly describe how we 1) detect and 2) investigate bugs in \vampire{}.
The main point of this section is that these two steps can be equally difficult.
The search space for \vampire{} is vast, and finding the combination of inputs that triggers a bug is very difficult. Some bugs are incredibly subtle, particularly soundness bugs or those involving memory errors, and tracking them down can involve hunting through thousands of lines of output.

\subsection{The Input Search Space}

The two inputs to \vampire{} are the input problem and a strategy capturing proof search parameters. 


The currently used proof search parameters in \vampire{} are summarised in Figure~\ref{fig:params}. We do not have the space (or desire) to describe these all here.\footnote{Explanations for most of these options can be obtained by running {\tt vampire -explain <option>}.} More than half of these options have more than two possible values and some can take arbitrary numbers. As an illustration of the size of this search space, if each option had just two values each and we could check each option combination in one second and had a time machine that could take us back to beginning of the world (4.5 billion years) we would only have covered about 0.0004\% of the possible combinations. 
Of course, this example is somewhat of an overestimate as there are parameter dependencies that shrink the search space and many parameter combinations will result in the same actual proof search. Nonetheless, the parameter search space is prohibitively large to explore systematically.

\begin{figure}[t]
\tiny{
\begin{multicols}{3}
{\tt age\_weight\_ratio }\\
{\tt avatar }\\
{\tt avatar\_add\_complementary }\\
{\tt avatar\_congruence\_closure }\\
{\tt avatar\_delete\_deactivated }\\
{\tt avatar\_eager\_removal }\\
{\tt avatar\_fast\_restart }\\
{\tt avatar\_flush\_period }\\
{\tt avatar\_flush\_quotient }\\
{\tt avatar\_minimize\_model }\\
{\tt avatar\_nonsplittable\_components }\\
{\tt backward\_demodulation }\\
{\tt backward\_subsumption }\\
{\tt backward\_subsumption\_resolution }\\
{\tt binary\_resolution }\\
{\tt block\_clause\_elimination }\\
{\tt cc\_unsat\_cores }\\
{\tt condensation }\\
{\tt demodulation\_redundancy\_check }\\
{\tt equality\_proxy }\\
{\tt equality\_resolution\_with\_deletion }\\
{\tt equational\_tautology\_removal }\\
{\tt extensionality\_resolution }\\
{\tt fmb\_adjust\_sorts }\\
{\tt fmb\_detect\_sort\_bounds }\\
{\tt fmb\_enumeration\_strategy }\\
{\tt fmb\_size\_weight\_ratio }\\
{\tt fmb\_symmetry\_ratio }\\
{\tt fool\_paramodulation }\\
{\tt forward\_demodulation }\\
{\tt forward\_literal\_rewriting }\\
{\tt forward\_subsumption }\\
{\tt forward\_subsumption\_resolution }\\
{\tt function\_definition\_elimination }\\
{\tt general\_splitting }\\
{\tt global\_subsumption }\\
{\tt increased\_numeral\_weight }\\
{\tt inequality\_splitting }\\
{\tt inline\_let }\\
{\tt inst\_gen\_big\_restart\_ratio }\\
{\tt inst\_gen\_passive\_reactivation }\\
{\tt inst\_gen\_resolution\_ratio }\\
{\tt inst\_gen\_restart\_period }\\
{\tt inst\_gen\_restart\_period\_quotient }\\
{\tt inst\_gen\_selection }\\
{\tt inst\_gen\_with\_resolution }\\
{\tt instantiation }\\
{\tt literal\_comparison\_mode }\\
{\tt literal\_maximality\_aftercheck }\\
{\tt lookahaed\_delay }\\
{\tt lrs\_weight\_limit\_only }\\
{\tt naming }\\
{\tt nongoal\_weight\_coefficient }\\
{\tt nonliterals\_in\_clause\_weight }\\
{\tt sat\_solver }\\
{\tt saturation\_algorithm }\\
{\tt selection }\\
{\tt simulated\_time\_limit }\\
{\tt sine\_depth }\\
{\tt sine\_generality\_threshold }\\
{\tt sine\_selection }\\
{\tt sine\_tolerance }\\
{\tt sos }\\
{\tt split\_at\_activation }\\
{\tt superposition\_from\_variables }\\
{\tt symbol\_precedence }\\
{\tt term\_algebra\_acyclicity }\\
{\tt term\_algebra\_rules }\\
{\tt theory\_axioms }\\
{\tt theory\_flattening }\\
{\tt time\_limit }\\
{\tt unit\_resulting\_resolution }\\
{\tt unused\_predicate\_definition\_removal }\\
{\tt use\_dismatching }\\
\end{multicols}}
\caption{The proof search parameters for \vampire{}\label{fig:params}.}
\end{figure}

\subsection{The Debug Process}

Given the large parameter space our current approach is to randomly sample the parameter space, under the additional dependency constraints, and sets of available problems. We use a cluster that enables us to carry out around a million checks a day (using varying short time limits). Once an error is detected, we must diagnose and fix the fault. Below we describe some of our methods for doing this.

\paragraph{Tracing.} Vampire has its own library for tracing function calls. A macro is manually inserted at the start of each significant function. This macro enables the tracing library to maintain the current call stack so that it can be printed on an assertion violation or during signal handling. Along with the call stack trace, the number of such call points passed so far will be returned. This then allows the developer to explicitly output the function calls leading up to the erroneous point by providing this value to the tracing library. This feature is invaluable in quickly locating the cause of an assertion violation and some cases of program crashes.

\paragraph{Memory Checking.} Vampire implements its own memory management library, allowing fine-grained control of memory allocation and deallocation. This also allows Vampire to enforce soft memory limits.
In debug mode, Vampire will keep track of each allocated piece of memory and check that the corresponding deallocation is as expected. Vampire will also report memory leaks by reporting on unallocated memory at the end of the proof search. 

\paragraph{Segmentation Faults and Silent Memory Issues.} The most difficult bug to debug is a rogue pointer or piece of uninitialised memory. Often such bugs are only noticed via incorrect results and much manual effort. We find that a first step of applying Valgrind\footnote{\url{http://valgrind.org}} will often detect the more straightforward issues.

\paragraph{Proof Checking.} To detect unsoundness we employ proof checking, which we discuss further in Section~\ref{sec:proofcheck}. We do not currently have a corresponding method for checking that a saturated set complies with necessary completeness conditions.

%% file: experience.tex
\section{A Sample of Bugs}\label{sec:bugs}

We now illustrate the kinds of bugs that can appear in \vampire{}. The majority of these bugs were detected during development, but still managed to exist in the development version of \vampire{} for some time before they were detected and fixed. We attempt to include explanations of why the bugs were not detected immediately, which informs our later discussion of what could be done better.

\subsection{A Very Effective Skolemisation Optimisation}

Skolemisation is a necessary step of the process used to translate an input formula into a clause (see \cite{GCAI2016:New_Techniques_in_Clausal_Form_Generation} for how \vampire{} implements this process). The standard transformation is as follows 
\[
(\exists x) (\varphi[x]) \quad \longrightarrow \quad \varphi[f_x(y_1,...,y_k)] 
\]
where $y_1,...,y_k$ are variables universally quantified in the containing formula. An optimised version of this transformation can be produced by noticing that in
\[
(\forall x) ((\exists y)(p(y,y)) \vee (\exists z)(q(x,z)))
\]
the variable $y$ does not rely on $x$ and $(\forall x)$ can be pushed in; this is called miniscoping. A buggy implementation of this optimisation introduced the transformation
\[
(\exists x) (\varphi[x]) \quad \longrightarrow \quad \varphi[f_x(y_1,...,y_k)] 
\]
where $y_1,...,y_k$ are variables universally quantified in the containing formula \emph{and} occurring in $\varphi[x]$. To understand why this is buggy consider
\[
(\forall u)(\exists x) (p(x,u) \land (\exists y) q(x,y))
\]
here we have $(\exists y) q(x,y)$ not containing $u$ so it would be Skolemised to $q(f(u),g)$ according to the above rule. 
However, this is incorrect as $x$ is itself dependent on the universally quantified $u$. The correct rule should instead be
\[
(\exists x) (\varphi[x]) \quad \longrightarrow \quad \varphi[f_x(y_1,...,y_k)] 
\]
where $y_1,...,y_k$ are variables universally quantified in the containing formula \emph{and} occurring in $\varphi[x] \sigma$ where $\sigma$ is a substitution containing previous Skolemisations.

With this corrected rule, the Skolemised example formula becomes: 
\[
p(f(u),u) \land q(f(u),g(u)).
\]

We expected this optimisation to improve performance, so when it did we did not immediately realise that this was due to \vampire{} solving a different, often slightly easier, problem. This was only detected when inspecting a different proof to understand a separate issue.

\subsection{Troubles in Parsing}

When the parser in \vampire{} was first extended to handle the unary minus arithmetic operator it erroneously parsed $-t$ as $(t)-(-t)$,
effectively resulting in $2t$. This was caused by incorrectly modifying a function that previously handled binary operators only. Whilst we would expect such an error to lead to a crash, the function instead fell through a case statement treated it as binary minus with $-t$ as the second term. This was not immediately detected as we did not have many non-theorem problems containing unary minus in our test set. 

\subsection{Memory Issues}

Bugs involving memory allocation can be very difficult to debug. In one case a class for representing propositional clauses declared an array
\begin{verbatim}
SATLiteral _literals[1];
\end{verbatim}
i.e. with implicit size of $1$ but then ensured that the correct amount of memory was allocated and initialised. With one exception: in the case of the empty clause no memory was allocated. Later the class was extended with extra fields and a non-trivial constructor, which implicitly initialised the array. In the case of the empty clause this caused a constructor to be called on an unallocated piece of memory i.e. a random piece of memory, most likely already used, was written to, leading to non-deterministically unsound behaviour. 

In a similar case, an implementation of skip lists employed a similar trick which behaved as expected in debug mode. However, in release mode a higher level of optimisation ({\tt -0 3}) was applied, which removed necessary code. It was unclear whether the trick or the optimisation were at fault but in either case the bug was difficult to diagnose as it only occurred in release mode.

\subsection{Problems with Hashing over Raw Data}

Vampire makes heavy use of data structures which rely on hashing.
For each class of objects that could be a key we typically have overloaded functions specifying what it means to hash such objects.
However, there is also a fallback implementation which simply hashes 
the \texttt{sizeof(o)}-many bytes starting from the address \texttt{\&o}.

This became the source of a hard to discover bug when moving to a new platform.
To meet a memory alignment requirement, a new compiler decided that a struct 
holding an \texttt{int} and a pointer should occupy 16 bytes,
but only 12 of these corresponded to the actually stored values.
The remaining 4 bytes of padding would in principle hold 
arbitrary values, while still participating in the hash computation.

Here the original code worked correctly and it was moving to a new platform that introduced the bug. Therefore, it was difficult to apply the \emph{what changed recently} question in a specific way.

\subsection{Inconsistent Theory Axioms}

For reasoning in theories \vampire{} adds \emph{theory axioms} (formulas such as $x+y=y+x$). When these were added the incorrect axiom
\[
0 \le x \vee {\sf abs}(-x) = x
\]
was added instead of
\[
\neg(0 \le x) \vee {\sf abs}(-x) = x
\]
where ${\sf abs}$ is meant to give the absolute value of a number. This allowed \vampire{} to derive an inconsistency using the added theory axioms alone. However, this was not detected for some time as the theory axiom was used to describe a feature 
that was rarely used and never used in a non-theorem problem in our testing set of problems. 
At another point a more subtle inconsistency was introduced that survived for a while as the proof of inconsistency from theory axioms was very long. Since encountering these issues we have added an assertion that a proof should contain formulas from the input!


\subsection{Misusing Z3}

In a recently added feature we make use of the Z3 SMT solver \cite{Z3}, which we use via its API. A number of bugs occurred due to misusing this API and Z3 in general. The first bug was based on failing to guard statements that could represent division by zero. 
Z3 treats division as an underspecified 
function and is allowed to assign any value to $t/0$.
This was inconsistent with our use of Z3 and led to unsound inference steps.
In another case we made use of an API call that returned an object without increasing a reference counter,
and consequently had memory issues as the object was sometimes deleted. A final bug was traced to Z3 itself and quickly fixed by their developers. This demonstrates the additional issues involved with integrating other tools. 



%
%

%% file: proofcheck.tex
\section{Proof Checking} \label{sec:proofcheck}

The easiest way to confirm a result indicating that the input formula is a theorem is to check that the associated proof only performs sound inference steps. This process is called proof checking and here we briefly describe 
the capabilities and limitations of the proof checking technique as currently realised in \vampire{}.

\subsection{Checking a Proof}

We introduce the idea of proof checking using an example (see our previous work \cite{Vampire2016:Better_Proof_Output_for_Vampire} for more information about proofs in \vampire{}). Given the clauses
\[
p(a) \quad\quad \neg p(x) \vee b=x \quad\quad \neg p(b)
\]
\vampire{} will produce the following proof

\begin{verbatim}
1. p(a) [input]
2. ~p(X0) | b = X0 [input]
3. ~p(b) [input]
4. a = b [resolution 2,1]
5. ~p(a) [backward demodulation 4,3]
7. $false [subsumption resolution 5,1]
\end{verbatim}

A proof is a directed acyclic graph printed in a linear form where nodes that have no incoming edges are either input formulas or axioms introduced by \vampire{}, and the single node with no outgoing edges contains the contradiction. In the above proof each derived clause is labelled with the name of the inference and the lines of the premises.

To check a proof we just need to establish that for each inference that its consequence logically follows from its premises.
By running {\tt vampire -p proofcheck} we can produce the following output in TPTP format which captures the three problems that need to be solved to check that the proof is correct.
\begin{verbatim}
fof(r4,conjecture, a = b ). %resolution
fof(pr2,axiom, ( ! [X0] : (~p(X0) | b = X0) ) ).
fof(pr1,axiom, p(a) ).
%#
fof(r5,conjecture, ~p(a) ). %backward demodulation
fof(pr4,axiom, a = b ).
fof(pr3,axiom, ~p(b) ).
%#
fof(r7,conjecture, $false ). %subsumption resolution
fof(pr5,axiom, ~p(a) ).
fof(pr1,axiom, p(a) ).
\end{verbatim}
We can pass these directly to an independent theorem prover and if a step cannot be independently verified then it should be investigated. 

\subsection{What We are Missing} \label{proofcheck:miss}

The above description suggests that we have a good method for checking the correctness of proofs. However, there are two problems with the above approach. The first problem is that in our experience it is not uncommon for a proof step to not be independently verified whilst still being sound. Such false positives take a lot of time to investigate. The second, and more substantial, problem is that there are parts of the proof process that cannot be handled by the above approach. There are two main classes of inferences that cannot be handled in this way and we describe them below:

\paragraph{Symbol Introducing Preprocessing.} Certain inference steps of the clausification phase, 
e.g. Skolemization and formula naming \cite{GCAI2016:New_Techniques_in_Clausal_Form_Generation}, 
introduce new symbols and as such do not preserve logical equivalence. This means the conclusion
of the inference does not logically follow from its premises.
What these steps only preserve is global satisfiability of the clause set they modify.
One necessary condition for correctness is that the introduced symbols be \emph{fresh},
i.e. not appearing elsewhere in the input. This cannot (in principle) be
checked by the described approach and requires a non-trivial extension.

\paragraph{SAT and SMT solving.} \vampire{} makes use of SAT and SMT solvers in various ways (see \cite{Vampire2014and2015:The_Uses_of_SAT_Solvers_in_Vampire}). This means that we have some inferences in \vampire{} that are of the form \emph{$P_1 \wedge \ldots \wedge P_n \rightarrow C$ by SAT/SMT}, or even the argument that some abstraction or grounding of the premises leads to $C$ by SAT or SMT solving. To handle such proof steps we need to collect together the premises (potentially apply the necessary abstraction or grounding) and run a SAT or SMT solver as appropriate.

%% file: challenges.tex
\section{Challenges} \label{sec:challenges}

We now present the main contribution of this paper, a discussion of what we have identified as the main challenges left to be solved, or at least addressed. These challenges are given in order of importance, as we perceive it.

\subsection{Full and Automated Proof Checking}

As described in Section~\ref{sec:proofcheck}, there is already reasonable support for independent checking the correctness of proofs. However, this situation could still be greatly improved.

\paragraph{Missing Features.} As outlined in Section~\ref{proofcheck:miss} there are parts of proofs that cannot currently be proof checked. To support checking these features might require adding additional information to the proofs.

\paragraph{Automating Proof Checking.} Having independent tools able to check the correctness of proofs is irrelevant if those tools are not used.
Ideally, theorem provers should provide the functionality to check the proofs that they produce automatically. In general, the problems produced during the proof checking process are easy to solve. Therefore, one could imagine a situation where, in a certain mode, a theorem prover created these problems and called an independent solver.

\paragraph{Independence.} In some cases it might not be possible to find an independent solver 
able to handle the problems produced by proof checking.
The solver might not be able to check an individual step, because it is too hard,
or not be able to handle the language features the problem contains (in the case of preprocessing steps).
In the former case, a weaker independence could be achieved by making use of a previous version of the original theorem prover that we are more confident in.

\subsection{Analysability of Unsound Proofs}

Checking whether a proof is correct or not is essential. However, knowing that a proof is incorrect is not, in itself, very useful. Another missing piece to this puzzle are tools that can analyse proofs and extract, summarise or explain the \emph{reason} the proof is incorrect. The proof checking process will reveal the proof step that fails to hold, but the problem of detecting the underlying reason for that proof step to have occurred is non-trivial. 

\subsection{Checking Non-theorem Results}

So far we have completely ignored the incorrect result of reporting a problem to be satisfiable when it is not. It is not clear how to practically check whether a saturated set is indeed saturated as the notion of saturation is dependent on used calculus and its instantiation with parameters
such as the term ordering and literal selection methods. One must also check that proof search never deletes anything that is not redundant.

Note that this problem is significantly more complex than proof checking. In proof checking we must check that each inference of the proof is sound i.e. that we were allowed to perform the inferences that we performed to derive a contradiction. If we have a saturated set then we should check that every inference that Vampire chose not to perform was redundant; this is what we often have to do manually, with some intuition about what such inferences might be. The number of such inferences is typically a few orders of magnitude larger than the length of a typical proof.

%

\subsection{Achieving Better Coverage with Random Testing}

As previously discussed, due to the enormous variability in proof search parameters and possible problem inputs, the best approach to detecting errors and incorrect results is through random search. However, the current approaches to random search are not optimal. Here we briefly outline areas of improvement. 

\paragraph{Code Coverage.} Our current approach makes no attempts to ensure that testing covers all lines in the code. Even though this is a very weak notion of coverage, it could be used to detect areas of code that should be tested, or removed if never used.

\paragraph{Coverage of the Parameter Space.} Whilst random sampling of the parameter space as done in Section~\ref{sec:random} can be effective at discovering bugs, it is not clear that all areas of the parameter space are of equal interest. Clearly, combinations of features that have not been tested together should have priority, and features added more recently should be tested more thoroughly.

\paragraph{Coverage of the Problem Space.} This is an area where relatively little has been done. Currently we use libraries of existing problems (such as TPTP \cite{TPTP} and SMTLIB \cite{BarST-SMTLIB}) as possible inputs to the testing process. However, as we saw in Section~\ref{sec:bugs}, if we do not have a problem that exercises a certain feature sufficiently, we are unlikely to detect bugs related to that feature. 
For example, the TPTP language contains features that are very rarely, sometimes never, used within the TPTP library. 
This issue is not confined to language features. Proof search is dependent on particular dimensions of the input problem (e.g. size, signature) that are difficult to quantify. If the input problems do not cover these dimensions sufficiently then certain parts of \vampire{} will not be tested effectively. This suggests that a useful area of research would be the automatic generation of problems, or \emph{fuzzing} of existing problems, to cover such dimensions. 


%% file: conclusion.tex
\section{Conclusion} \label{sec:conclusion}

This paper describes our experience testing the Vampire theorem prover and what we see as the challenges to overcome to help us improve this effort. The ideas we discuss here generalise to other theorem provers and some efforts, such as good proof checking techniques and better problem coverage, would be widely beneficial.

Addressing the challenges set out in this paper is part of our current research and we are currently working on providing a publicly available proof checking tool that can fully and automatically check proofs produced by Vampire.